\documentclass[12pt,a4paper,final]{iopart}

\usepackage[inner= 2cm, outer=2cm, top= 2.5 cm,bottom= 2.5 cm]{geometry} 
\usepackage{iopams}  
\usepackage{graphicx}
\usepackage[colorlinks=true,linkcolor=blue,urlcolor=blue,citecolor=blue]{hyperref}
\usepackage{multirow}
\usepackage{caption}
\usepackage{color}\clearpage\maketitle
\usepackage{float}
\usepackage{amsfonts}
\usepackage{cite}
\begin{document}
	
\title{A simple metamaterial absorber associated with Fano-like resonance}

\author{Raghwendra Kumar$^{a,b}$ and S. Anantha Ramakrishna$^{a}$}

\address{ $^{a}$ Department of Physics, Indian Institute of Technology Kanpur, Kanpur, 208016, India\\
	$^{b}$ Department of Physics, Bihar National College, Patna, 800004, India
	}

\ead{raghawk@iitk.ac.in}

\begin{abstract}	
A simple metamaterial absorber suitable for fabrication over large areas associated with a Fano-like resonance is proposed. The proposed designed of the metamaterial absorber consists of photoresist disk arrays on a silicon substrate followed by the deposition of a tri-layer of Au/ZnS/Au on the top of the structure. Due to the tri-layer, there is formation of a cavity between the substrate and gold layer on the top of photoresist disk and a waveguide in the ZnS layer in between the gold layers on the substrate and the ZnS layer itself. The Fano-like resonance arises due to the interference of the cavity mode and the guided-mode resonance as well as the Wood's anomaly. The cavity mode resonance works as continuum mode  whereas the guided mode resonance and Wood's anomaly work as discrete modes. The spectral position of the resonance can be tuned by just controlling the thickness of the tri-layer instead of the structural size and shape modification of the micro/nano-structures as usually done in conventional metamaterial absorbers. This design can be easily scaled for fabrication over large areas as it separate the structuring and deposition processes, and makes them sequential thereby avoiding expensive and complex lift-off or etching processes which are usually required for conventional metamaterial processing.		
\end{abstract}
\vspace{2pc}


\section{Introduction}

Electromagnetic (EM) metamaterials are artificial nano/micro-structured composites of sub-wavelength metallic and dielectric components that exhibit exotic electromagnetic responses such as complete absorption of electromagnetic waves~\cite{landy2008perfect,watts2012metamaterial, dayal2012design, guddala2015thermally, pradhan2017high}, negative refraction~\cite{shelby2001experimental}, invisibility cloaking~\cite{schurig2006metamaterial}, sub-diffraction limited imaging using super lenses~\cite{pendry2000negative}, etc, which are not available in natural materials~\cite{veselago1968electrodynamics}. These composites usually consist of identical structured unit cells, arranged typically in a periodic fashion, and can be considered as bulk homogeneous media described as an effective medium through effective medium parameters~\cite{ramakrishna2008physics}. Among these responses, the complete/perfect absorption of EM radiations as metamaterial absorbers have received considerable attention due their potential applications for radar stealth at microwave frequencies, sensitive infrared detectors and micro-bolometer arrays~\cite{ogawa2012wavelength}, controllable thermal emitters for manipulating the black body emittance~\cite{liu2011taming} and thermophotovoltaics~\cite{wu2012metamaterial} at infrared frequencies, etc. A perfect metamaterial absorber with nearly perfect absorption caused by simultaneous excitation of electric and magnetic resonances to create an impedance match of a resonant medium with the surrounding medium was first proposed by Landy et al. in 2008~\cite{landy2008perfect}. After that, a lot of absorbers based on different physical mechanisms have been demonstrated theoretically and experimentally in a wide spectral range~\cite{ameling2010cavity, hedayati2014review,rhee2014metamaterial}. It has been possible to easily adapt metamaterial for different frequencies due to the scale invariance of Maxwell's equations.  The main purpose of metamaterial absorber is to reduce undesired reflections and to increase absorption within the given spectral range.

Metamaterial absorbers can be realized using different physical mechanisms such as anti-reflection coating, resonant/impedance matching~\cite{mattiucci2013impedance, zhu2010dual,  tao2008metamaterial,avitzour2009wide}, localized surface plasmons resonance, Ohmic losses in the visible regime and lossy dielectric in the microwave regime, cavity resonance~\cite{bhattarai2017metamaterial}, guided mode resonance~\cite{zhou2013extraordinary}, etc. A typical metamaterial absorber has a tri-layered structure, with a structured top conducting layer separated by a continuous dielectric layer from a bottom conducting layer. The excitation depends on the electric resonance of the top layer and the impedance matching arises from the magnetic resonance created by the sheet current distributions on the top resonating structure and the bottom conducting layer (image charges and currents), which are anti-parallel. The thickness of the dielectric layer determines the capacitance for the LC resonance and the optimal impedance matching for coupling maximum radiation into the resonant structure. Generally, metamaterial absorbers show symmetric (Gaussian) absorption profiles, however the asymmetric absorption profiles have been also absorbed due to the interference of different modes arises in the system~\cite{pu2013investigation, le2015enhanced}.      \par

Usually, the Fano resonance~\cite{fano1961effects,miroshnichenko2010fano} is used to describe asymmetric resonance, which arises due to the constructive and destructive interference of discrete resonance states with broad band continuum states. This phenomenon and the underlying mechanism, being common and appearing in many domains of physical sciences, are found in a variety of micro/nano-structures such as plasmonics and metamaterials~\cite{luk2010fano,gallinet2011ab}, photonic crystals~\cite{galli2009light,rybin2009fano,zhou2014progress}, quantum dots~\cite{johnson2004coulomb}, etc. The asymmetric Fano resonance profile promises a variety of applications in a wide range of photonic devices, such as sensors~\cite{hao2008symmetry,lee2017highly}, optical filters~\cite{zhou2007fano}, switches~\cite{nozaki2013ultralow}, detectors~\cite{wu2012fano,zhang2014coherent}, non-linear devices~\cite{kroner2008nonlinear} and slow-light devices~\cite{wu2011broadband}, etc. In the last few years, with advancements in micro/nano fabrication techniques and the development of integration techniques, Fano resonance has been widely investigated in micro/nano fabricated structures~\cite{luk2010fano,gallinet2011ab,zhou2014progress,
	khanikaev2013fano,gu2012active,singh2009coupling}.\par

The metamaterial perfect absorbers  based on impedance matched resonances typically involve a complex design, and the fabrication tolerances are quite severe. However, for many infra-red applications, such as solar thermal devices or thermal camouflage, control of the emissivity and the absorptivity is required over very large area surfaces ranging from few square meters to even square kilo meters. It is a great manufacturing challenge to produce metamaterials that have complex micrometric or even sub-micrometric structures containing metallic as well as dielectric constituents over these large areas with good fidelity and over realistic time scales. In this work, we present numerical studies on a novel design of metamaterial absorber for controlling the emissivity/absorptivity at infra-red frequencies. The metamaterial consists of a tri-layer of Au/ZnS/Au thin films deposited on a structured photoresist disk arrays, and depends on a cavity and guided mode resonance (GMR) of the structure to produce high absorptivity. The simplified design of the metamaterial absorber can be easily implemented for sequential patterning and deposition processes without lift-off or etching, making the process applicable for large area metamaterial surfaces~\cite{kumar2018development}.
 
\section{Design of the metamaterial by electromagnetic simulations}

\begin{figure*}[ht]
\begin{center}
	\includegraphics[scale=0.45]{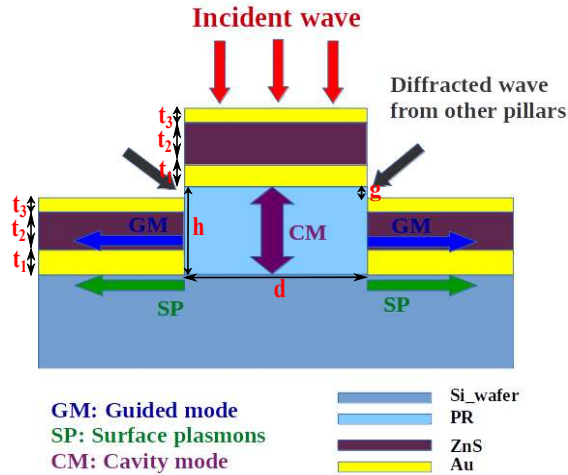}
	\caption{\label{Schematic}Schematic diagram of the unit cell of the metamaterial absorber with illustration of different existing modes. $t_{1}$, $t_{2}$ and $t_{3}$ are the thickness of the bottom gold layer, ZnS layer and top gold layer, respectively. $h$ and $d$ are the height and diameter of the photoresist disk. $g$ is the gap between gold layers on the top of photoresist disk and the ZnS layer on the silicon substrate.}
\end{center}
\end{figure*}

The unit cell of the metamaterial absorber is shown in Fig.~\ref{Schematic}. It consists of a photoresist disk of height ($h$) 500 nm and diameter ($d$) 1.68 $\mu$m on a silicon substrate with consecutive layers of bottom gold of thickness $t_{1}$ (160 nm), ZnS of thickness $t_{2}$ (300 nm) and top gold of thickness $t_{3}$ (30 nm) on the top of both the disk as well as on the other parts of the silicon substrate not covered by the disk. In essence, the structure consists of three disks, out of two are gold disks and one is ZnS disk placed on the photoresist disk, with continuous thin films of Au/ZnS/Au containing holes filled with the photoresist at a lower level on the substrate. Due to the tri-layer deposition, a cavity is formed between the bottom layer of gold on the top of photoresist disk and silicon substrate, because the refractive index of the photoresist is lower than that of silicon, and gold is itself highly reflecting. There is formation of waveguide in the ZnS layer between the gold layers on the silicon substrate and the ZnS itself.\par

The numerical simulation is performed by radio frequency module of COMSOL Multiphysics software~\cite{Comsol} based on the finite element method. Periodic boundary conditions were used along the transverse x and y directions to simulate an infinite array of unit cells to reduce the complexity of the simulation. Perfectly matched layers (PML) were used along the propagation direction (z - direction) to prevent the reflection of the waves from the top and bottom domain boundaries with sufficient distance from the structure. The dispersive refractive indices (real and imaginary parts) of evaporated gold, ZnS and silicon were taken from Refs~\cite{olmon2012optical, debenham1984refractive} and \cite{chandler2005high}, respectively. The refractive index of the photoresist in simulation is considered as 1.58. The transmittance ($T$) and reflectance ($R$) of the structure is calculated by using port conditions at a plane above the structure in air and below the structure in the silicon substrate. The absorbance is calculated using energy conservation as $A = 1- R -T$ at each wavelength. \par

\begin{figure}[ht]
\begin{center}
\includegraphics[scale=0.65]{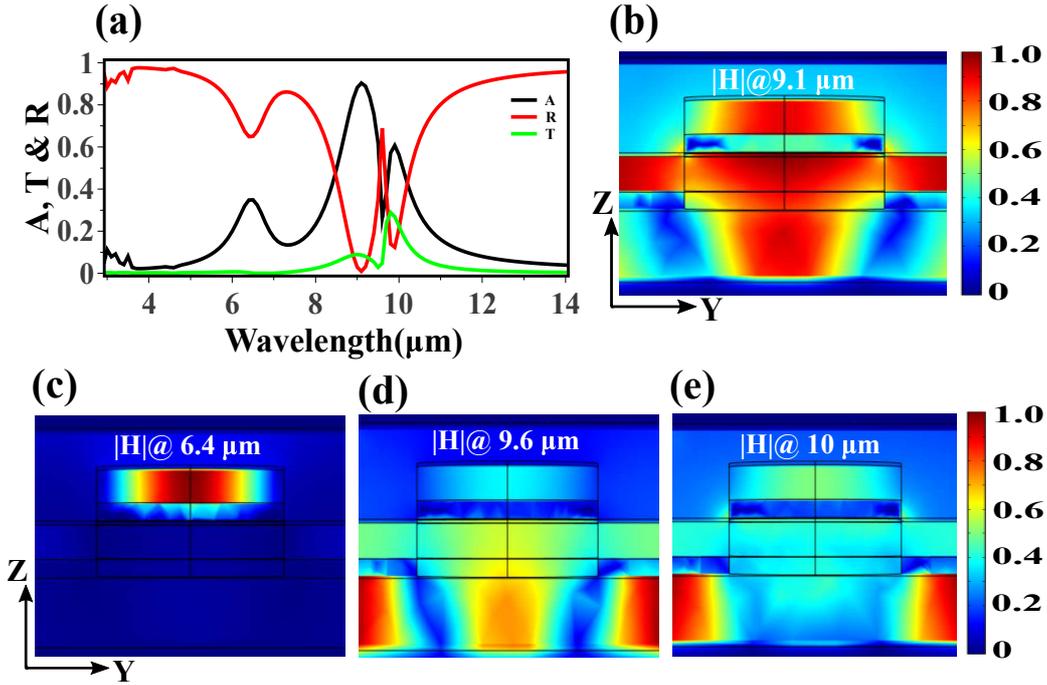}
\caption{\label{ART_normH}(a) simulated reflectance, transmittance and absorbance spectra corresponding to the metamaterial absorber, (b) color plot of the normalized magnetic field at resonant wavelength 9.1 $\mu $m, (c) color plot of the normalized magnetic field at resonant wavelength 6.4 $\mu $m, (d) color plot of the normalized magnetic field at the resonant wavelength 9.6 $\mu $m, (e) color plot of the normalized magnetic field at the resonant wavelength 10.0 $\mu $m. The parameters used for simulations: periodicity of the disk array, disk height, disk diameter, ZnS thickness, bottom layer Au thickness, top layer Au thickness are 2.8 $\mu $m, 0.5 $\mu $m, 1.68 $\mu $m, 300 nm and 30 nm, respectively.}
\end{center}
\end{figure}

The simulated reflectance, transmittance and absorbance spectra of the metamaterial are shown in Fig.~\ref{ART_normH}(a). The transmittance spectrum consists of two peaks at wavelengths 9.8 $\mu $m and 9 $\mu $m. The reflectance spectrum consists of three minima at wavelengths of 6.4 $\mu $m, 9.1 $\mu $m and 9.85$\mu $m. The absorbance spectrum consists of three maxima at wavelengths of 6.4 $\mu $m, 9.1 $\mu $m and 10 $\mu $m. Color maps of the normalized magnetic field at different resonant wavelengths 9.1 $\mu $m, 6.4  $\mu $m, 9.6  $\mu $m and 10 $\mu $m are shown in Figs.~\ref{ART_normH}(b), (c), (d) and (e), respectively. It is clear from Fig.~\ref{ART_normH}(b) that the fields are strongly confined in the waveguide as well as in the cavity at a wavelength of 9.1 $\mu $m. The field is highly localized in the ZnS layer between the two layers of gold on the top of photoresist disk at wavelength of 6.4 $\mu $m. Figs.~\ref{ART_normH}(d) and (e) show the normalized magnetic field maps at the peak and dip of the corresponding transmittance spectrum, respectively. It is clear from Fig.~\ref{ART_normH}(d) that the fields are distributed along the Au/Si interface and weakly within the cavity. The fields are distributed along the Au/Si interface and not localized within the cavity at 10 $\mu $m wavelength (see Fig.~\ref{ART_normH}(e)). It is clear that there is no contribution of the resonance at wavelength 6.4 $\mu $m in the resultant interfering resonances at 9.1 $\mu $m wavelength. In order to understand the contribution of the different resonances in the interference at 9.1 $\mu $m, we note that the resonance at 6.4 $\mu $m, which is not close to 9.1 $\mu $m can be eliminated by removing the top layer gold and ZnS from the photoresist disk.  \par

\begin{figure}[ht]
\begin{center}
\includegraphics[scale=0.7]{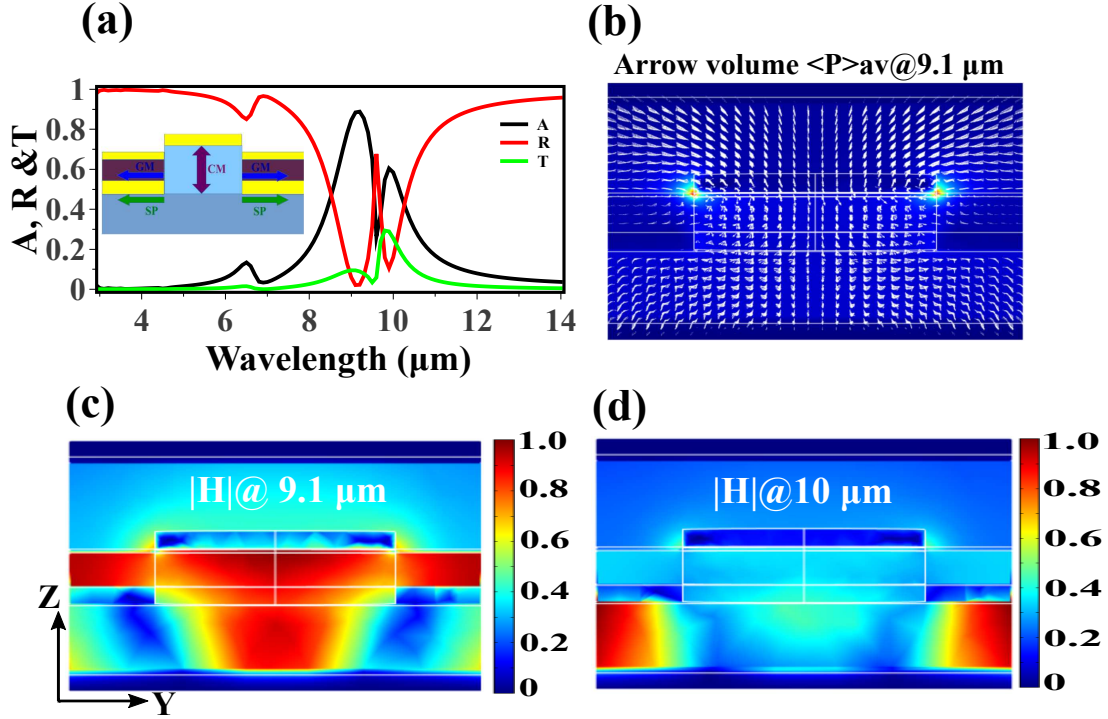}
\caption{\label{ART_normH_without_top}(a) simulated reflectance, transmittance and absorbance spectra corresponding to the metamaterial absorber without top Au and ZnS layers on the disk, (b) Color plot of the normalized time average power with Poynting vector field lines at resonant wavelength 9.1 $\mu $m, (d) color plot of the normalized magnetic field at the resonant wavelength 9.1 $\mu $m, (e) color plot of the normalized magnetic field at the resonant wavelength 10.0 $\mu $m. The parameters used for simulations: periodicity of the disk array, disk height, disk diameter, ZnS thickness, bottom gold layer thickness and top gold layer thickness are 2.8 $\mu $m, 0.5 $\mu $m, 1.68 $\mu $m, 300 nm, 160 nm and 30 nm, respectively.}
\end{center}
\end{figure}

 The simulated reflectance, transmittance and absorbance spectra of the metamaterial absorber without the ZnS and top Au layers on the photoresist disk are shown in Fig.~\ref{ART_normH_without_top}(a), where the corresponding unit cell shown in the inset. Fig.~\ref{ART_normH_without_top}(b) shows the normalized time average power flow in color map, and to understand the power flow in the system, an arrow plot of the time-averaged power is plotted in the same figure. It is clear that the power flows into the cavity and the waveguide as well as along the interface of gold and silicon. The color maps of the normalized magnetic fields at wavelengths of 9.1 $\mu $m and 10 $\mu $m are shown in Figs.~\ref{ART_normH_without_top}(c) and (d), respectively. 

\begin{figure*}[ht]
\begin{center}
\includegraphics[scale=0.7]{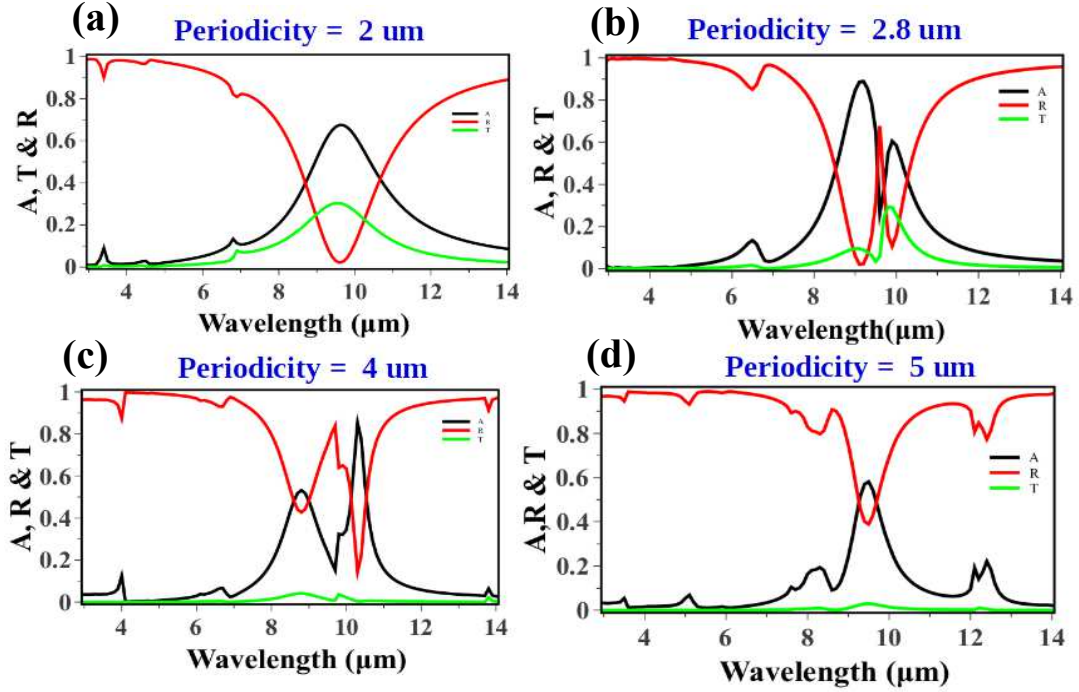}
\caption{\label{Periodicity}Simulated reflectance, transmittance and  absorbance spectra at normal incidence for different periodicities of the photoresist disk array. The parameters used for simulations: disk height, disk diameter, bottom Au layer thickness, ZnS layer thickness and top Au layer thickness are  0.5 $\mu $m, 1.68 $\mu $m, 160 nm, 300 nm and 30 nm, respectively.}
\end{center}
\end{figure*}

Fig.~\ref{Periodicity} shows the transmittance, reflectance and absorbance spectra for different periodicities of the photoresist disk array. The other parameters like the thickness of the bottom and top gold layer, ZnS layer and height of the photoresist disk are kept constant. With increasing the periodicity of the disk array, an extra resonance appears. The extra resonance position shifts towards longer wavelengths with increasing the periodicity of photoresist disk array. But, the resonance corresponding to the cavity mode remains almost independent of the periodicity of the disk array. There is an asymmetric resonance for a fixed value of periodicity 2.8 $\mu $m, which arises due to the interference of the fixed resonance (cavity resonance) and the resonance is caused by changing the periodicity (Guided-mode resonance) of the disk array [Fig.~\ref{Periodicity}(b)]. The interfering resonances corresponding to the different modes separate when the period of the disk array becomes 4 $\mu $m (see in Fig.~\ref{Periodicity}(c)). At the periodicity of 2 $\mu $m, only one resonance appears that arises due to the confinement of fields into the cavity, whereas at 5 $\mu $m periodicity an extra resonance appears around 12.4 $\mu $m. \par

\begin{figure*}[ht]
 \begin{center}
 \includegraphics[scale=0.75]{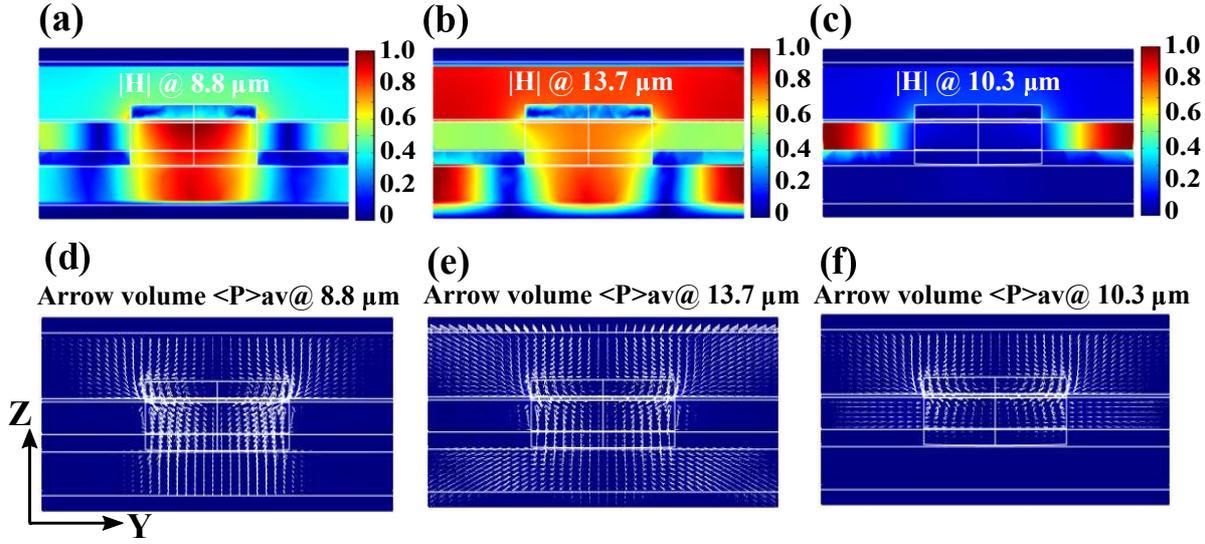}
 \caption{\label{normH_power}Color plots of the normalized magnetic field at resonance wavelength (a) $\lambda$ = 8.8 $ \mu $m, (b) $\lambda$ = 10.3 $ \mu $m and (c) $\lambda$ = 13.7 $ \mu $m. The arrow volume plots of the time average power flow at resonance wavelength (d) $\lambda$ = 8.8 $ \mu $m, (e) $\lambda$ = 10.3 $ \mu $m and (f) $\lambda$ = 13.7 $ \mu $m.}
 \end{center}
 \end{figure*}

Fig.~\ref{normH_power} shows the color-maps of the normalized magnetic field and arrow volume plot of the normalized time average power flow at different resonance wavelengths for 4 $\mu $m periodicity of the disk array. Fig.~\ref{normH_power}(a) shows the color plot of the normalized magnetic field at 8.8 $\mu $m wavelength, and at this resonance wavelength field is only localized in the cavity formed between gold layers on the photoresist disk and Si substrate. The corresponding color map of the normalized time average power and its arrow volume plot is shown in the figure~\ref{normH_power}(d). Fig.~\ref{normH_power}(b) shows the color map of the normalized magnetic field at 13.7 $\mu $m wavelength, and at this resonance wavelength field is only extended into the silicon substrate from the gold layer side. The corresponding color map of the normalized time average power and its arrow volume plot are shown in the figure.~\ref{normH_power}(e). The arrow power flow indicates the power flow in the substrates along the gold and silicon interface. Fig.~\ref{normH_power}(c) shows the color map of the normalized magnetic field at 10.3 $\mu $m wavelength and at this resonance wavelength field is localized in the sandwiched ZnS layer between the bottom gold layer on the  substrate and the top gold layer on ZnS. The corresponding color map of the normalized time average power and its arrow volume plot is shown in Fig.~\ref{normH_power}(f). It is clear that most of the power is flowing into the ZnS sandwiched layer.\par
 
\begin{figure}[h!]
\begin{center}
	\includegraphics[scale=0.85]{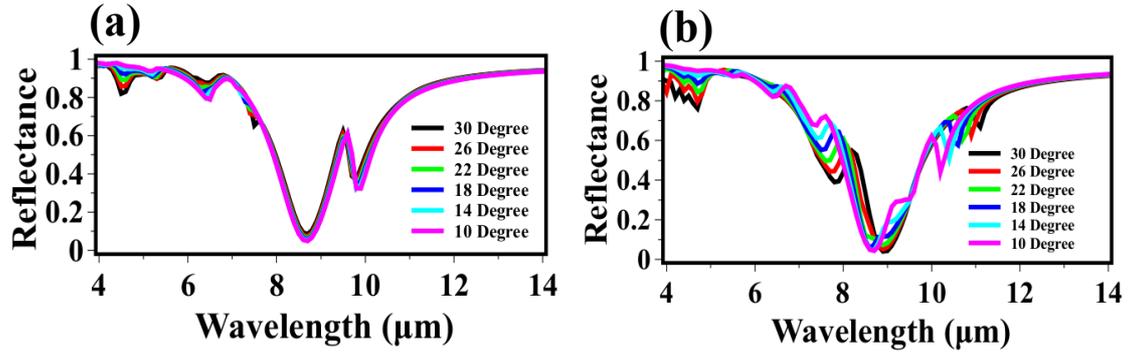}
	\caption{\label{TE_TM_Pol_sim}The simulated reflectance spectra of the fabricated sample for the (a) TE polarization, (b) TM polarization at different incident angles. The parameters used for simulations: disk height, disk diameter, bottom Au layer thickness, ZnS layer thickness and top Au layer thickness are  0.5 $\mu $m, 1.68 $\mu $m, 155 nm, 300 nm and 30 nm, respectively.}
\end{center}
\end{figure}

Figs.~\ref{TE_TM_Pol_sim}(a) and (b) show the simulated reflectance spectra at different incident angles for the TE and TM polarizations, respectively. There is a slight decrement in the minimum of reflectance spectra around the 9 $\mu$m wavelength with increasing the angle of incidence in the case of TE polarization, whereas it remains almost constant with a slight redshift in the case of TM polarization. The minimum at around 10 $ \mu $m wavelength in the case of TE polarization slightly decreases in amplitude with increasing the angle of incidence whereas it slightly decreases in amplitude with redshift in the case of TM polarization. An extra minimum appears in the reflectance spectra around  7.5 $ \mu $m wavelength in the case of TM polarization, and it shifts towards longer wavelengths with increasing the angle of incidence. The minimum in the reflectance spectra around  6.5 $ \mu $m wavelength remains almost constant with increasing the angle of incidence in both polarizations. The minimum in the reflectance spectrum around 4.5 $ \mu $m wavelength slightly increases in amplitude with increasing the angle of incidence for both polarizations but faster in the case of TM polarization compared to that of TE polarization. Splitting appears in the minimum at 4.5 $ \mu $m wavelength with increasing the angle of incidence in the case of TM polarization, whereas there is no such splitting in the case of TE polarization.

\begin{figure}[h!]
\begin{center}
	\includegraphics[scale=0.8]{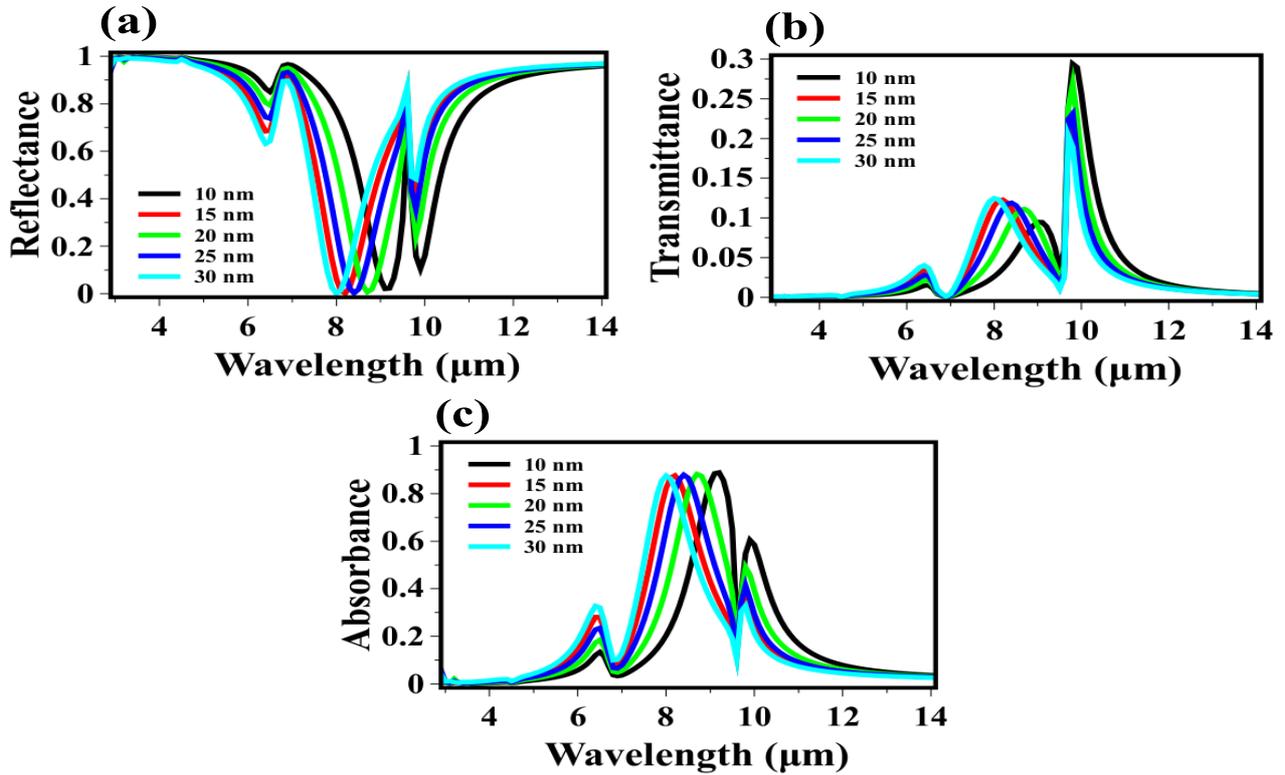}
	\caption{\label{Gap}The simulated (a) reflectance, (b) transmittance and (c) absorbance spectra of the tri-layered metamaterial for different values of $g$ as 10 nm, 15 nm, 20 nm, 25 nm and 30 nm at normal incidence. The other parameters such as disk height, disk diameter, ZnS layer thickness and top Au layer thickness are 0.5 $\mu $m, 1.68 $\mu $m, 300 nm and 30 nm, respectively.}
\end{center}
\end{figure}

Fig.~\ref{Gap} shows the simulated electromagnetic response of the tri-layered metamaterial absorber for different values of the $g$ parameter at normal incidence. The value of $g$ was changed by changing the thickness of the bottom gold layer, and the other parameters are kept constant. In the reflectance spectra [see Fig.~\ref{Gap}(a)], the minimum at 9.1 $\mu $m wavelength shifts towards shorter wavelengths, and the minimum at 9.8 $\mu $m wavelength remains almost constant with increasing the value of $g$. While the minimum at 6.4 $\mu $m wavelength increases with increasing the gap. In the transmittance spectra [see Fig.~\ref{Gap}(b)], the fundamental order transmission amplitude at 9.8 $\mu $m wavelength decreases whereas the higher order transmission amplitude at 6.4 $\mu $m wavelength increases with increasing the gap. The transmission peak at 9.1 $\mu $m wavelength shifted towards shorter wavelengths with increasing the gap. In the absorbance spectra [see Fig.~\ref{Gap}(c)], the absorption peak around the 10 $\mu $m wavelength decreases with a minimal shift towards shorter wavelengths with increasing the gap. There is a shift in the absorption peak around 9 $\mu $m towards shorter wavelengths with increasing the gap whereas the absorption peak around 6.4 $\mu $m wavelength increases in amplitude with increasing the gap.\par

\begin{figure*}[ht]
\begin{center}
	\includegraphics[scale=0.8]{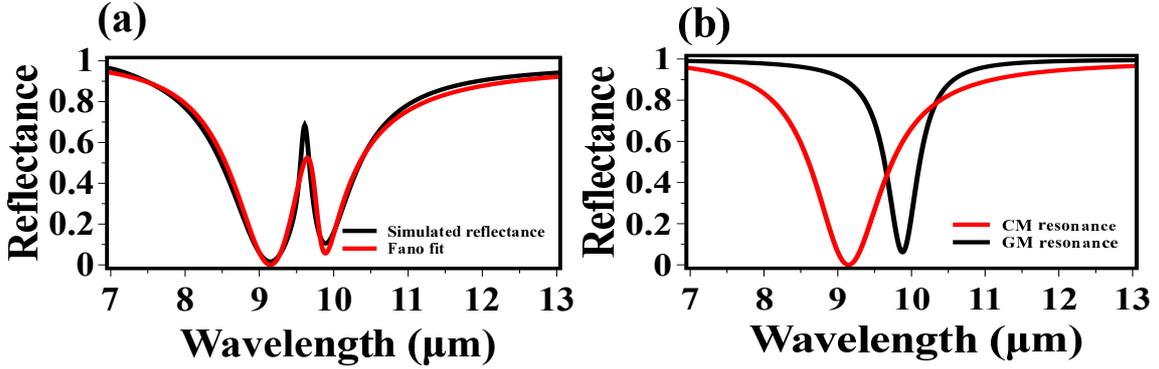}
	\caption{\label{Fano_fit_and_indi}(a) Simulated reflectance spectrum (black line) with Fano fit (red line), (b) Fano fitted reflectance spectra of the corresponding individual resonance. The parameters used for simulations: periodicity of the disk array, disk height, disk diameter, bottom layer Au thickness, ZnS layer thickness and top layer Au thickness are 2.8 $\mu $m, 0.5 $\mu $m, 1.68 $\mu $m, 160 nm, 300 nm and 30 nm, respectively. The fitting parameters; $\omega_{1} = 65.6$ THz, $\omega_{2} = 60.73$ THz, $\gamma_{1} = 6.42$ THz, $\gamma_{2} = 1.86$ THz, $\gamma^{'}_{1} = 6.4$ THz, $\gamma^{'}_{2} = 3.0$ THz and $\omega_{0} = 6.53$ THz.}		
\end{center}
\end{figure*}

Fig.~\ref{Fano_fit_and_indi}(a) shows the simulated reflectance spectrum (black line) at normal incidence of the metamaterial absorber with the Fano fit (red line) using the coupled mode theory of one port and two modes. There are two minima at 9.1 $\mu $m and 9.8 $\mu $m wavelengths in the reflectance spectrum, which corresponds to the cavity mode resonance and guided mode resonance, respectively. Fig.~\ref{Fano_fit_and_indi}(b) shows a fit of the Fano line shape combining the individual resonance. This was plotted using the same fitting parameters as used to fit the asymmetric reflectance spectrum in Fig.~\ref{Fano_fit_and_indi}(a). The individual response was plotted by switching off the other corresponding resonance. For example for CM resonance, the value of $\omega_{2} = 0$, and similarly for GMR, the value of $\omega_{1} = 0$ [see Eq.~\ref{Fit}]. 

\section{Discussion}\label{Disc}
The mechanism of absorption of EM waves can be understood from the illustration of different excited modes, shown  in the schematic diagram of the unit cell of the metamaterial absorber [see Fig.~\ref{Schematic}]. Formation of a cavity between the bottom layer gold on the top of photoresist disk and silicon substrate occurs because of the lower refractive index of the photoresist than that of silicon. The diffracted wave from the disk gets multiply reflected between the gold layer and the silicon substrate and causes the trapping of the electromagnetic wave into the photoresist medium~\cite{artar2009fabry,lee2006design}. There is a formation of a  waveguide in the ZnS layer sandwiched between the gold layer the on the silicon substrate and ZnS layer itself. Waves diffracted from the different disks enter through the gap between the bottom layer of gold on the photoresist disk and top layer gold on the ZnS layer on uncovered portions of the substrate by the photoresist disk. The diffracted light, which enters through the gap couples into the cavity mode as well as to the waveguide mode~\cite{rosenblatt1997resonant}, and some parts of it just propagate at the interface of silicon and gold layers. The same happens by coupling through the gap with other disks. Thus, diffracted coupled waves travel in forward and backward directions in the waveguide and cause the formation of standing waves. The diffracted waves propagating along the plane at the interface between gold and silicon get absorbed, and do not contribute to the transmittance, and the corresponding phenomenon is known as Wood's anomaly~\cite{maystre2012theory, hessel1965new}.\par

Fig.~\ref{ART_normH}(a) shows the simulated electromagnetic response of the metamaterial absorber. The asymmetric electromagnetic response  is caused by  simultaneous excitation and interaction of the cavity mode, guided mode and Wood's anomaly resonances, which results in a Fano-like resonance. This Fano-like resonance arises due to the interference of the continuum mode (cavity mode resonance) and the dark mode (guided mode resonance and  Wood's anomaly) resonances. The simulated absorption spectrum consists of three peaks at wavelengths 6.4 $\mu $m, 9.1 $\mu $m and 10 $\mu $m. The peak at a wavelength of  6.4 $\mu $m corresponds to the metamaterial absorption peak due to the presence of the tri-layer (Au/ZnS/Au) on the top of the photoresist disk~\cite{landy2008perfect,cui2014plasmonic, dayal2012design}. There is no contribution of this resonance in the Fano-like resonance, so it was neglected in the further simulations by not including the top gold and ZnS layer on the photoresist disk in the simulations.

Fig.~\ref{ART_normH_without_top} shows the simulated electromagnetic response without considering the top gold and ZnS layer on the photoresist disk. However, there is still an appearance of a peak in the absorption spectrum at 6.4 $\mu $m which is due to the localized surface plasmons at the corner of the gold disk placed on the top of the photoresist disk. There is no effect of the removal of the top gold and ZnS layers from the photoresist disk on the other resonances, which is seen in Fig.~\ref{ART_normH_without_top}(a). The arrow plot of the normalized time average power flow with its color map is shown in  Fig.~\ref{ART_normH_without_top}(b). The simultaneous excitation of cavity mode, guided mode and Wood's anomaly can be seen through the coupling of power flow into each mode. The simultaneous excitation of all these modes can be seen in the color map of the normalized magnetic fields in Fig.~\ref{ART_normH_without_top}(c). The propagation of the diffracted wave at the interface of gold and silicon can be seen in  Fig.~\ref{ART_normH_without_top}(d), which results in almost zero transmittance. \par

To understand the Fano-like resonance better and to elucidate the contribution of different modes in the Fano resonance, the simulations were carried out for different periodicities of the photoresist disk array which are shown in Fig.~\ref{Periodicity}. Only the cavity mode resonance exists in the case of photoresist disk array with 2 $\mu $m periodicity. The existence of this mode is independent of the periodicity of the disk array and is always associated with the unit cell, and hence regarded as continuum mode, whereas the existence of the guided mode resonance and Wood's anomaly is conditional and hence regarded as the discrete mode. Interference of the continuum and discrete modes is conditional, and it only happens at 2.8 $\mu $m periodicity (Fig.~\ref{Periodicity}b). For periodicity of 4 $\mu $m, these interfering modes get separated in frequency (Fig.~\ref{Periodicity}c), and the modes corresponding to the resonance wavelengths of 8.8 $\mu $m, 10.3 $\mu $m and 13.7 $\mu $m are the cavity mode, guided mode and Wood's anomaly, respectively. The existence of the cavity mode can be seen even at 5 $\mu $m periodicity (Fig.~\ref{Periodicity}d). 

The color maps of the normalized magnetic field and normalized time average power flow with its arrow volume plot at different resonant wavelengths corresponding to 4 $\mu $m periodicity of disk array are shown in Fig.~\ref{normH_power}. Excitation of cavity mode resonance can be understood by the confinement of the magnetic field in the cavity and arrow volume plot of the time average power from Figs.~\ref{normH_power}(a) and (d), respectively. Excitation of the guided mode resonance can be understood by the confinement of the magnetic field in the waveguide and arrow volume plot of the time average power from Figs.~\ref{normH_power}(c) and (f), respectively. Similarly, excitation of the Wood's anomaly resonance can be understood from Figs.~\ref{normH_power}(d) and (e).\par

The effect of spoof SPPs can be understood from the angle dependent reflectance spectra of the metamaterial absorber for the TE and TM polarizations [see Figs.~\ref{TE_TM_Pol_sim}(a) and (b)]. In case of TE polarization, magnitude of electric remains constant so there is only slight variation in the magnitude of deeps and peaks in the reflectance spectra at different incident angles. In case of TM polarization, magnitude of electric field changes due to which coupling condition for spoof SPPs changes, and hence there is finite variation in the magnitude of deeps and peaks as well as spectral shift in the resonance~\cite{kumar2018enhanced}.\par

The minimum in the reflectance spectra as well as the corresponding peak in absorbance and transmittance spectra around 9 $\mu $m wavelength shift towards shorter wavelengths with increasing the value of $g$ [ Fig.~\ref{Gap}]. This can be understood from the circuit model theory of metamaterial absorber. The resonance frequency can be defined as $\omega = 1/\sqrt LC $ on the basis of the circuit model, where $L$ and $C$ are effective inductance and capacitance corresponding to the metamaterial absorber. In the proposed metamterial absorber with increasing the value of $g$ there is decreased in the effective capacitance which corresponds to increase in the resonance frequency and hence shorter wavelengths shift in resonance.  

The coupled mode theory (CMT) has also been used to describe the Fano asymmetric line profile which arises due to the interference of continuum and discrete mode resonances in the metamaterial absorbers~\cite{khanikaev2013fano,fan2003temporal}. Since our system  contains two main resonances; one is cavity resonance and other is GMR, so the CMT with the one-port and two modes resonator model can be used to describe the asymmetric reflectance spectra of the metamaterial absorber ~\cite{suh2004temporal,zhou2014progress}. The reflectance {\it r} corresponding our metamaterial absorber is  
\begin{equation}\label{Fit}
r = -1 + \frac{2(W_{1}\gamma_{2} + W_{2}\gamma_{1}-2\gamma_{0}\sqrt{\gamma_{1}\gamma_{2}})}{(W_{1}W_{2}-|\gamma_{0}|^2)}
\end{equation}
where $ W_{i} = j(\omega - \omega_{i} + \gamma_{i} + \gamma^{'}_{i}), (i = 1, 2)$, $\omega_{i} $ is resonant frequency corresponding to the $i^{th}$ mode, $\gamma_{i}$ is the radiation loss rate due to the interaction between the $i^{th}$ mode and the outgoing wave, $\gamma^{'}_{i}$ is the intrinsic loss rate associated with the $i^{th}$ mode, and $\gamma_{0}$ is the coupling between the modes. The first two terms of the numerator of the second part of Eq.~\ref{Fit} describe the responses of two modes independently after coupling modulation, whereas the last term $\sqrt{\gamma_{1}\gamma_{2}}$ describes the responses of the cross-coupling between the two modes after coupling modulation. For the metamaterial absorber, the radiation losses for the two modes are $\gamma_{1} = 6.42$ THz and $\gamma_{2} = 1.86$ THz. The intrinsic loss rates of the two modes are $\gamma^{'}_{1} = 6.4$ THz  and $\gamma^{'}_{2} = 3.0$ THz. The coupling between modes occurs at $\omega_{0} = 6.53$ THz. The resonant frequencies of the two modes are $\omega_{1} = 65.6$ THz and $\omega_{2} = 60.73$ THz. 

The fitting parameters match well with the expected parameters such as the spectral position of resonances, bandwidth of the individual resonance, etc. Now, it is clear that the asymmetric profile is caused by the interference of mainly the CM resonance and the GM resonance. The response of the individual resonance using the same fitting parameters as for the fitting of the reflectance spectrum [see in Fig.~\ref{Fano_fit_and_indi}(a)] is plotted in Fig.~\ref{Fano_fit_and_indi}(b). 

\section{Conclusions}\label{Cons}
We have proposed a very simple design of a metamaterial absorber with band selective absorptivity/emissivity that is suitable for large area fabrication. The origin of absorption as well as the asymmetric nature of the absorption profile has been understood. The absorption is caused due to the simultaneous excitation of the cavity and guided mode resonances in the structures whereas the asymmetric profile is caused due to the interference of the continuum mode, discrete modes (guided mode and Wood's anomalies). The advantage of this design is that it can be easily implemented for fabrication over large areas by separating the structuring and deposition processes and making them sequential, as well as avoiding expensive and complex lift-off or etching processes. Our approach can be easily adopted for rapid productions of large areas fabrication on the industrial level. The spectral position of the resonance can be tuned without structural size and shape modification just by changing the gap between the gold layers on the top of photoresist pillar and the ZnS layer at the substrate. 
 
\section{Acknowledgement}
RK thanks Council of Scientific $\& $ Industrial Research (CSIR), India for fellowship.  
\section*{References}
\bibliographystyle{apsrev}
\bibliography{mybibfile1}
\end{document}